\documentclass[10pt,journal,compsoc]{IEEEtran}
\ifCLASSOPTIONcompsoc
  \usepackage[nocompress]{cite}
\else
  \usepackage{cite}
\fi
\ifCLASSINFOpdf
\else
\fi

\usepackage{stfloats}
\fnbelowfloat
\usepackage{booktabs} 
\usepackage{multirow}
\usepackage{listings}
\usepackage{graphicx}
\usepackage{epsfig}
\usepackage{xcolor}
\usepackage{amsmath}
\usepackage{tabularx}
\usepackage{hyperref}
\usepackage[capitalise]{cleveref} 
\usepackage{relsize}
\usepackage{xspace}
\usepackage{array}

\def\|#1|{\mathid{#1}}
\newcommand{\mathid}[1]{\ensuremath{\mathit{#1}}}
\def\<#1>{\codeid{#1}}
\protected\def\codeid#1{\ifmmode{\mbox{\smaller\ttfamily{#1}}}\else{\smaller\ttfamily #1}\fi}

\newcommand{\smalltitlecolon}[1]{{\smallskip \noindent \bf  {#1}:\ }}

\newcommand{\yxmodifyok}[2]{#2}

\usepackage{listings}

\begin{document}
%
\title{An Empirical Study of Fault Localization Families and Their Combinations}
%
%
%
%

\author{Daming~Zou,
        Jingjing~Liang,
        Yingfei~Xiong,
        Michael~D.~Ernst,
        and Lu~Zhang

\IEEEcompsocitemizethanks{
\IEEEcompsocthanksitem D. Zou, J. Liang, Y. Xiong, and L. Zhang 
are with the School of
Electronics Engineering and Computer Science, Institute of Software,
Peking University, Beijing 100871, PR China, and the
Key Laboratory of High Confidence Software Technologies, Ministry of Education,
PR China.\protect\\
Email: \{zoudm, jingjingliang, xiongyf, zhanglucs\}@pku.edu.cn.
\IEEEcompsocthanksitem M.D. Ernst is with the Department of
Computer Science \& Engineering, University of Washington, Seattle, WA
98195.\protect\\
E-mail: mernst@cs.washington.edu.
\IEEEcompsocthanksitem Yingfei Xiong is the corresponding author.
}
}

%
%

\markboth{}{}
\IEEEtitleabstractindextext{%
\begin{abstract}
  The performance of fault localization techniques is critical to
  their adoption in practice. This paper reports on an empirical study
  of a wide range of fault localization techniques on real-world
  faults. Different from previous studies, this paper (1) considers a
  wide range of techniques from different families, (2)
  combines different techniques, and (3) considers the execution time of different techniques. Our results
  reveal that a combined technique significantly outperforms any
  individual technique (200\% increase in faults localized in Top 1),
  suggesting that combination may be a desirable way to apply fault
  localization techniques and that future techniques should also be
  evaluated in the combined setting. Our
  implementation is publicly available for evaluating and combining
  fault localization techniques. 
\end{abstract}

}

\maketitle

\IEEEdisplaynontitleabstractindextext

%
\IEEEpeerreviewmaketitle

\IEEEraisesectionheading{\section{Introduction}\label{sec:introduction}}


\IEEEPARstart{T}{he} goal of fault localization is to
identify defective program elements related to software failures.
Automated fault localization
uses static
and run-time information about the
program to identify program
elements that may be the root cause of the failure.
This paper considers seven families of fault localization techniques, which take as input seven different types of information:

{\hbadness=10000
  \begin{itemize}
  \item  Spectrum-based fault localization
    (SBFL)~\cite{xie2013theoretical,abreu2007accuracy,harrold2000empirical}:
    utilizing test coverage information
  \item  Mutation-based fault localization
    (MBFL)~\cite{papadakis2015metallaxis,moon2014ask}: utilizing test
    results from mutating the program
  \item  Dynamic program
    slicing~\cite{agrawal1995fault,renieres2003fault}: utilizing
    dynamic program dependencies
  \item  Stack trace analysis~\cite{wong2014boosting,wu2014crashlocator}:
    utilizing crash reports
  \item  Predicate
    switching~\cite{Zhang2006Locating}: utilizing
    test results from mutating the results of
    conditional expressions
  \item  Information-retrieval-based fault localization (IR-based FL)~\cite{zhou2012should}: utilizing bug report information
  \item  History-based fault
    localization~\cite{kim2007predicting,rahman2011bugcache}:
    utilizing the development history
\end{itemize}
}


Some techniques compute a suspiciousness score for each program element
and can generate a ranked list of elements,
such as spectrum-based fault localization. Other techniques
only mark a set of elements as suspicious, such as dynamic program
slicing.

\yxmodifyok{Fault localization techniques are evaluated based on where in their list the
real fault appears.
Fault localization techniques are helpful to programmers only when the
programmer sees the root cause occur at a high absolute rank in the list~\cite{parnin2011automated}.
In a survey, programmers stated that they believe a successful fault
localization technique should locate the fault in the top 5 positions, but
this belief has not been empirically evaluated~\cite{kochhar2016practitioners}.
Most empirical studies to
date~\cite{pearson2017evaluating,jones2005empirical,abreu2009practical,harrold1998empirical}
have evaluated techniques that use test
information, i.e., SBFL and MBFL\@.
The performance of other techniques on real-world defects is not
known.
}
{
  The performance of fault
  localization is critical to its adoption in practice. Fault localization
  techniques are helpful only when the root causes are ranked at a
  high absolute position~\cite{parnin2011automated,xia2016automated}, such as
  within the top 5~\cite{kochhar2016practitioners}.
  A number of empirical
  studies~\cite{pearson2017evaluating,jones2005empirical,abreu2009practical,harrold1998empirical}
  have evaluated the performance of SBFL and MBFL\@. However, no
  empirical study has evaluated the performance of
  other techniques on real-world faults, as far as we know.
}


This paper reports on an empirical study of
a wide range of fault localization techniques from different families.
Following the insight from
existing work~\cite{pearson2017evaluating} that the performance of fault
localization techniques may differ between real faults and artificial
faults, our study is based on 357 real-world faults from the Defects4J
dataset~\cite{just2014defects4j}.

Our study has two main novel aspects. First, since
techniques in different families use different
information sources, it is interesting to know how much these
techniques are correlated to each other. We measured the correlation
between different pairs of techniques and explored the possibility of combining
these techniques using the learning to rank model~\cite{burges2005learning}. In
contrast, previous work usually considers techniques in one or a few families~\cite{wang2017lightweight}, e.g., combining different formulae in
SBFL~\cite{xuan2014learning} or combining SBFL and history-based techniques~\cite{sohn2017fluccs}, and our work, CombineFL,
 is the first to explore
combinations of a wide range of techniques that rely on different information sources.

The second novelty is that we measured the time cost of different fault localization
techniques. Existing studies have shown that efficiency and scalability are both critical to the adoption of fault localization techniques~\cite{kochhar2016practitioners}. Thus, a good fault localization approach must balance between localization performance and cost. We have considered different usage scenarios to find the best balance in practice.

We also improved the measurement of fault localization performance by designing a new measurement $E_{\|inspect|}$ that calculates the expected rank when multiple faulty elements are presented in ties.

Finally, we have released our experimental infrastructure CombineFL-core and
the fault localization data of the studied techniques,
 which can be used by other researchers to
evaluate fault localization techniques and to combine different fault
localization techniques.

Our study has the following main findings:
\begin{itemize}
  \item On real-world faults, 
  all techniques except for Bugspots and BugLocator localize more than 6\% of faults in the top 10.
  The best family, SBFL, localizes about 44\% faults of in the top 10.
 
  \item Most techniques in our study are weakly correlated with each another, especially those in different families, indicating
  the potential of combining them.
 
  \item CombineFL improves performance significantly: 200/63/51/31\%
    increase in localized faults in the top 1/3/5/10,
    compared to the best standalone technique.

  \item CombineFL also outperforms the four state-of-the-art fault localization approaches, MULTRIC~\cite{xuan2014learning}, Savant~\cite{b2016learning}, FLUCCS~\cite{sohn2017fluccs}, and TraPT~\cite{li2017transforming} by 133\%, 167\%, 11\% and 18\% in Top 1 correspondingly.
  \item Time costs of different fault localization families can be
    categorized into several levels. When using a technique at one time
    cost level, it does not affect run time to include all techniques from the preceding
    levels, but it does improve fault
    localization effectiveness.
  \item The above findings hold at both statement and method granularities
    --- that is, when the FL technique is identifying suspicious statements
    and when it is identifying suspicious methods.
  \end{itemize}

To sum up, the paper makes the following contributions.
\begin{itemize}
\item The first empirical study that compares a wide range of
  fault localization techniques on real faults.
\item A combined technique, CombineFL, which is configurable based on the time cost, and the peak performance of the technique significantly outperforms standalone techniques.
\item An infrastructure, CombineFL-core, for evaluating and combining fault
  localization techniques for future research.
\end{itemize}

The rest of the paper is organized as follows.
\Cref{sec:bg} presents background about several fault localization families.
\Cref{sec:exp-setup} gives the empirical evaluation methodology.
\Cref{sec:exp-result} shows the experiment results and answers the research questions.
\Cref{sec:related} discusses related research.
\Cref{sec:implication} discusses the implications for future research.
\cref{sec:conclusion} concludes.




\section{Background}\label{sec:bg}

Commonly, a fault localization technique takes as input a faulty program
and a set of test cases with at least one failed test, and it generates as output a
potentially ranked list of suspicious program elements. Recently, some approaches~\cite{zhou2012should,wen2016locus,rahman2011bugcache} considered other input information, such as the bug report or the development history. This paper also considers these approaches.
The common levels of
granularity for program elements are statements, methods, and files. This
paper uses statements as program elements, except for
\cref{sec:method-level,sec:compare-other-l2r} which compare results for different granularities.

This section first introduces seven families of fault localization
techniques, and then introduces the learning to rank model for combining
different techniques.

\subsection{Spectrum-Based Fault Localization}\label{sec:bg-sbfl}


A program spectrum is a measurement of run-time behavior, such as code
coverage~\cite{harrold2000empirical}. Collofello and Cousins proposed that
program spectra be used for fault
localization~\cite{collofello1986towards}.  Comparing program spectra on
passed and failed test cases enable ranking of program elements.  The more
frequently an element is executed in failed tests, and the less frequently
it is executed in passed tests, the more suspicious the element is.

Typically, an SBFL approach calculates suspiciousness scores using 
a ranking
metric~\cite{naish2011model,hao2005similarity,hao2005eliminating}, or risk evaluation
formula~\cite{xie2013theoretical,yoo2012evolving}, based on four values
collected from the executions of the tests, as shown in 
\Cref{tab:notation-spectrum}. For example,
Ochiai~\cite{abreu2007accuracy} and DStar~\cite{wong2014dstar} are effective SBFL techniques~\cite{yoo2012evolving,le2015should,pearson2017evaluating} using the formulas: 

\begin{eqnarray*}
\|Ochiai|(\|element|) & = & \frac{e_f}{\sqrt{(e_f+n_f)\cdot(e_f+e_p)}}
\\
\|DStar|(\|element|) & = & \frac{e_f^{*}}{e_p+n_f}
\end{eqnarray*}

\noindent
DStar's notation `*' is a variable, which we set to 2 based on the
recommendation from Wong et al.~\cite{wong2014dstar}.


\begin{table}
  \centering
  \caption{Input Values for Spectrum-Based Fault Localization}
  \label{tab:notation-spectrum}
  \setlength{\tabcolsep}{.6\tabcolsep}
  \begin{tabularx}{\columnwidth}{ll}
  \toprule
  $e_f$ & Number of failed tests that execute the program element. \\
  $e_p$ & Number of passed tests that execute the program element. \\
  $n_f$ & Number of failed tests that do not execute the program element. \\
  $n_p$ & Number of passed tests that do not execute the program element. \hfill\\
  \bottomrule
  \end{tabularx}
\end{table}

\subsection{Mutation-Based Fault Localization}\label{sec:bg-mbfl}
Mutation-based fault localization uses information from mutation
analysis~\cite{jia2011analysis},
rather than from regular program execution, as inputs to its
ranking metric or risk evaluation formula.
While SBFL techniques consider whether a statement is executed or not,
MBFL techniques consider whether the execution of a statement affects the
result of a test by injecting mutants. A mutant typically changes one
expression or statement by replacing one operand or expression with
another~\cite{pearson2017evaluating}. If a program statement
affects failed tests more frequently and affects passed tests more rarely,
it is more suspicious.

For a statement $s$, a MBFL
technique:
\begin{itemize}
\item generates a set of mutants $m(s) = \langle m_1(s), m_2(s), ... \rangle$, 
\item assigns each mutant a score $S_{\|mutant|}(m_i(s))$, and
\item aggregates the scores to a statement suspiciousness score $S_{\|statement|}(s)$.
\end{itemize}

MUSE~\cite{moon2014ask} and Metallaxis-FL~\cite{papadakis2015metallaxis}
are two state-of-the-art MBFL techniques.

MUSE assigns each mutant a suspiciousness score as follows:
$$ S_{\|mutant|}(m_i) = \|failed|(m_i) - \frac{f2p}{p2f} \cdot \|passed|(m_i)$$
where $\|failed|(m_i)$ is the number of test cases that failed on the original
program but now pass on a mutant $m_i$, and likewise for
$\|passed|(m_i)$. $f2p$ is the number of test cases that change
from fail to pass on any mutant, and likewise for $p2f$. To
aggregate mutant suspiciousness scores into a statement suspiciousness score, MUSE
uses $S_{\|statement|}(s) = Avg_{m_i \in m(s)}S_{\|mutant|}(m_i)$.

Metallaxis assigns each mutant a suspiciousness score using the Ochiai formula:
$$ S_{\|mutant|}(m_i) = \frac{\|failed|(m_i)}{\sqrt{\|totalfailed| \cdot (\|failed|(m_i)+\|passed|(m_i))}}$$
where $\|failed|(m_i)$ is the number of test cases that failed on the
original program and now the output changes on a mutant $m_i$, and
similarly for $\|passed|(m_i)$. $\|totalfailed|$ is the total number of
test cases that fail on the original program.

A mutant is said to be \emph{killed} by a test case if the test case has
different execution results on the mutated program and the original
program~\cite{zhang2016predictive}. A test case that kills mutants may
carry diagnostic information.
Note that the definition of
\emph{killed} in MUSE and Metallaxis is different. In MUSE, a failed test
case must change to passed to count as killing a mutant. In Metallaxis, a
failed test case only needs to generate a different output (may still be
failed) to count as killing a mutant.


\subsection{Program Slicing}\label{sec:bg-slicing}
A slicing criterion is a set of variables at a program location; for
example, they might be variables that have unexpected or undesired values.
A program slice is a subset of program elements that potentially affect
the slicing criterion~\cite{xu2005brief}.

Program slicing was introduced
as a debugging tool to reduce a program to a minimal form while still
maintaining a given behavior~\cite{weiser1981program}.  Static slicing
only uses the source code and accounts for all possible executions of the
program.

Dynamic slicing focuses on one execution for a specific
input~\cite{korel1988dynamic}.
The key difference between dynamic slicing and
static slicing is that dynamic slicing only includes executed statements
for the specific input, but static slicing includes possibly-executed statements for
all potential inputs. Since dynamic slices are significantly
smaller,
they are more suitable and effective for program debugging~\cite{zhang2007study}.

The following example shows the difference between static slicing and dynamic slicing.

\begin{lstlisting}[language=C,
    numberstyle=\small\ttfamily,
    basicstyle=\small\ttfamily]
int collatz(int x) {
1:  int res;
2:  if ((x % 2) == 0)
3:      res = x / 2;
4:  else
5:      res = x * 3 + 1;
6:  return res;
}
\end{lstlisting}

The \<collatz> function returns $x/2$ when $x$ is even and returns $3x+1$
when $x$ is odd. Let the slicing criterion be \<res> at line 6.  Static
slicing includes statements in both the \<then> and \<else> block, because
both may affect the value of \<res>.  Dynamic slicing considers a
particular execution of the program. For example, for \<x>=3, the dynamic
slice would contain line 5, but would not contain line 3.

\subsection{Stack trace Analysis}\label{sec:bg-stack}
A stack trace is the list of active stack frames during execution of a
program. Each stack frame corresponds to a function call that has not yet
returned. Stack traces are useful information sources for developers
during debugging tasks. When the system crashes, the stack trace
indicates the currently active function calls and the point where the crash
occurred.



\subsection{Predicate Switching}
Predicate switching~\cite{Zhang2006Locating} is a fault localization
technique
designed for faults related to control flow.
A predicate, or conditional expression, controls the execution
of different branches. If a failed test case can be changed to a passed test
case by modifying the evaluated result of a predicate, the predicate is
called a critical predicate and may be 
the root cause of the fault.

The technique first
traces the execution of the failed test and identifies all instances of branch
predicates. Then it repeatedly re-runs the test, forcibly switching the
outcome of a different predicate each time.
If switching a
predicate produces the correct output, the predicate is potentially the cause of
the fault and is called a critical predicate. 

Predicate switching is similar to MBFL techniques, as they both apply
mutations and examine the change of the execution results. We treat
predicate switching as a different family because predicate switching
mutates the control flow
rather than the program itself. For example, if a conditional expression
has been evaluated multiple times during the program execution, predicate
switching inverses one evaluation at a time instead of all
evaluations. Furthermore, previous
work~\cite{pearson2017evaluating,li2017transforming} does not include
predicate switching as an MBFL approach as far as we are aware.

\subsection{Information Retrieval-Based Fault Localization}

Information Retrieval (IR) was initially used to index text and
search for documents~\cite{Baeza-Yates:2011:MIR:1796408}. Recent
studies~\cite{zhou2012should,saha2013improving,wen2016locus,Zhang2015} have
applied information retrieval techniques to fault localization. These
approaches take as input a bug report, rather than a set of test cases, and
generate as output a list of relevant source code
files~\cite{wong2016survey}.

These approaches treat the bug reports as a query and then rank the source code files by their relevance to the query. Unlike aforementioned fault localization families, IR-based fault localization techniques do not require program execution information, such as passed and failed test cases. They locate relevant files based on the bug report~\cite{zhou2012should}.


\subsection{History-Based Fault Localization}
Program files that contained more
bugs in the past are likely to have more bugs in the
future~\cite{moser2008comparative}.
Development history can be used for \emph{fault prediction}, which
ranks the elements in a program
by their likelihood to be defective.
Traditionally, fault prediction and fault localization are
considered as different problems, and fault prediction runs before any failure has been
discovered~\cite{kim2007predicting}. However, since they both produce a
list of suspicious elements, this paper also considers fault prediction
techniques.

We consider a simple fault prediction technique
introduced by Rahman et al.~\cite{rahman2011bugcache}. This technique
ranks files by the number of fixing changes applied on them.
This simple technique has the same utility for inspections as
a more sophisticated fault prediction technique,
FixCache~\cite{kim2007predicting}.



\subsection{Learning to Rank}
Learning to rank techniques train a machine learning model for a ranking
task~\cite{li2011short}. Learning to rank is widely used in Information
Retrieval (IR) and Natural Language Processing
(NLP)~\cite{liu2009learning}. For example, in document retrieval, the task
is to sort documents by their relevance to a query.
One way to create the ranking model is with expert knowledge.
By contrast, learning
to rank techniques improve ranking performance and automatically create the
ranking model, integrating many features (or signals).

Liu categorized learning to rank models into three 
groups~\cite{liu2009learning}. Pointwise techniques transform the rank problem
into a regression or ordinal classification problem for the ordinal score in the
training data. Pairwise techniques approximate the problem by a classification
problem: creating a classifier for classifying item pairs according to their
ordinal position. The goal of pairwise techniques is to minimize ordinal
inversions. Listwise techniques take ranking lists as input and evaluate the
ranking lists directly by the loss functions.

Recently, Xuan and Monperrus showed that learning to rank
model can be used to combine different formulae in SBFL~\cite{xuan2014learning}. The basic idea is to treat the
suspiciousness score produced by different formulae as features and use
learning to rank to find a model that ranks the faulty element as high
as possible. In this paper we apply learning to rank similarly to
combine techniques from different families.




\section{Experimental Methodology}\label{sec:exp-setup}

\subsection{Experiment Overview and Research Questions}\label{sec:exp-overview}
Our experiments investigate the following research questions.

\smalltitlecolon{RQ1} How effective are the standalone fault localization techniques?

This question helps us to understand the performance of widely-used techniques.

\smalltitlecolon{RQ2} Are these techniques correlated?
What is their correlation?

This question explores the possibility of combining different
techniques. If different techniques are not correlated, then combining
them may archive better performance.

\smalltitlecolon{RQ3} How effectively can we combine these techniques
using learning to rank?

This question considers a specific way of combining different
techniques, and evaluates the performance of the combined technique. 

\smalltitlecolon{RQ4} What is the run-time cost of standalone
techniques and combined techniques?


The previous question only concerns the effectiveness of the
combined techniques. This question considers the efficiency. The
best technique for a given use case balances effectiveness and efficiency.

\smalltitlecolon{RQ5} Are the results the same for statement and method
granularity?

{
We shall answer the preceding four questions first at statement
granularity, which is often used in evaluating fault localization
approaches~\cite{xie2013provably,yoo2012evolving,hao2006towards,xie2014cooperative} and in
downstream applications such as program repair~\cite{xiong2017precise,Jiang:2018:SPR:3213846.3213871,xiong2018identifying}.
On the other hand, several studies
have suggested that methods may be a better granularity for
developers~\cite{b2016learning,poshyvanyk2007feature}, so we repeated the
experiments for the above questions at the method granularity.
}

\smalltitlecolon{RQ6} How effective the combined approach is when compared with the state-of-the-art techniques?

Recently, a set of new fault localization approaches were
proposed. Interestingly, they also use learning to rank to combine existing
techniques or other features. This research question compares the performance
of our combined approach to these approaches.

\subsection{Experimental Subjects}

Our experiments evaluate fault localization techniques on the
Defects4J~\cite{just2014defects4j} benchmark, version v1.0.1
(\cref{tab:dataset}). Defects4J contains 357 faults minimized
from real-world faults in five open-source Java projects. Many previous studies
on fault localization used Defects4J as their
benchmarks~\cite{ma2015grt, b2016learning, pearson2017evaluating}. For
each fault, Defects4J provides a faulty version of the project, a
fixed version of the project, and a suite of test cases that contains
at least one failed test case that triggers the fault.

\begin{table}[tbp]
  \centering
  \caption{Defects4J Dataset (version 1.0.1).
  `Faults' is the number of defective versions of the program.  `LoC' is
    average lines of code for each buggy version of the project, as reported by
    cloc\protect\footnotemark.
}
  \label{tab:dataset}
  \begin{tabular}{l|rr}
    \toprule
    Project & Faults & LoC\\
    \midrule
    Apache Commons \textbf{Math} & 106 & 103.9k\\
    Apache Commons \textbf{Lang} & 65 & 49.9k\\
    Joda-\textbf{Time} & 27 & 105.2k\\
    JFree\textbf{Chart} & 26 & 132.2k\\
    Google \textbf{Closure} compiler & 133 & 216.2k\\
    \midrule
    Total & 357 & 138.0k\\
    \bottomrule
  \end{tabular}
\end{table}

\footnotetext{\url{https://github.com/AlDanial/cloc}}

\subsection{Studied Techniques and Their Implementations}

A fault localization technique outputs one of the following:
\begin{itemize}
  \item A ranked list. Examples include the SBFL, MBFL, stack trace, and
    history-based families, and one slicing technique.
  \item A suspicious set. The techniques do not distinguish the
    suspiciousness between these elements.  Examples include the predicate
    switching family and some slicing techniques.
\end{itemize}

\subsubsection{SBFL and MBFL}
Pearson et al.~\cite{pearson2017evaluating} studied the performance of SBFL and
MBFL on Defects4j, and our experiments reuse their infrastructure and the
collected test coverage information.
For SBFL, we used only
the two techniques that performed best in Pearson et al.'s study:
Ochiai~\cite{abreu2007accuracy} and DStar~\cite{wong2014dstar}. The
parameter $*$ in DStar is set to $2$.
For MBFL, we used the two mainstream
MBFL techniques, MUSE~\cite{moon2014ask} and
Metallaxis~\cite{papadakis2015metallaxis}.
The formulae for calculating the suspiciousness are introduced
in \cref{sec:bg-sbfl,sec:bg-mbfl}.

\subsubsection{Dynamic Slicing}\label{sec:impl-slicing}

The JavaSlicer dynamic slicing
tool~\cite{hammacher2008bachthesis}
is based on the dynamic slicing algorithm of Wang and
Roychoudhury~\cite{wang2008dynamic,wang2004using}, with extensions for
object-oriented programs.  The {JavaSlicer} implementation attaches to the
program as a Java agent and rewrites classes as they are loaded into the
Java VM\@.


A test fails by throwing an exception, either because of a violated
assertion or a run-time crash. If there is only a single failed test, we use
the execution of the statement that throws the exception as the slicing
criterion.
The slice then contains
all statements that may have affected the statement that throws the exception.

If there are multiple failed tests,
our experiments apply three
strategies from a previous study~\cite{pan1992heuristics} to utilize multiple
slices: union,  intersection, and frequency. The first two strategies
calculate the union or the intersection
of the slices and report a set of statements as results. The
frequency strategy calculates the inclusion frequency for each
statement and reports a ranked list of statements based on the
frequency. The more frequently a statement is included in the slice of
a failed test, the more suspicious the statement is.\looseness=-1

\subsubsection{Stack Trace Analysis}\label{sec:exp-setup-stack}


According to Schroter et al.~\cite{schroter2010stack},
if the stack trace includes the faulty method
around 40\% of the
faults can be located in the very first frame, and 90\% of the faults
can be located within the top 10 stack frames.



We defined a stack trace technique based on this insight.
If the exception is thrown by the testing framework (such as JUnit),
then the technique returns an empty suspicious list.
Otherwise, we call the fault a crash fault and 
the suspicious list consists of the frames
in the stack trace. 
The frame at depth $d$ is given suspiciousness score $1/d$ score.
The score of an element (method) is its maximum score in all failed tests.\looseness=-1

\subsubsection{Predicate Switching}


We re-implemented Zhang et al.'s method of
predicate switching for Java (the original implementation was for
x86/x64 Linux binaries)~\cite{Zhang2006Locating}. Our implementation
of the technique is based on Eclipse Java development tools (JDT\@).
The technique first traces the execution of the failed test case and
records all executed predicates. Then it forcibly switches the
outcome of a predicate at run time. Once switching a predicate makes
the failed test case pass, it reports the predicate as a critical
predicate. The technique produces a set of critical predicates as the
suspicious program elements.




\subsubsection{IR-based Fault Localization}

BugLocator~\cite{zhou2012should} ranks all files based on the
textual similarity between the initial bug report and the source code file
using a revised Vector Space Model (rSVM\@).

\label{file-suspicousness-to-statment-suspiciousness}
Since the granularity in BugLocator is source file, our
implementation
uses the file score for all statements in
it. For example, if BugLocator reports that \textit{file1.java} has
suspiciousness score 0.2, then it marks every executable statement in
\textit{file1.java} with suspiciousness score 0.2.

\subsubsection{History-Based Fault Localization}
Bugspots\footnote{\url{https://github.com/igrigorik/bugspots}} is an implementation of
Rahman
et al.'s algorithm~\cite{rahman2011bugcache}.
Bugspots collects
revision control changes with
descriptions related to \textit{`fix'} or \textit{`close'}. The
tool ranks more recent bug-fixing changes higher
than older ones.

the granularity in Bugspots is source file. Our implementation maps the
score of a suspicious file to all executable statements as in
\cref{file-suspicousness-to-statment-suspiciousness}.

Bugspots supports only \textit{Git}
repositories. However, the version control system of \textit{Chart} in
Defects4J uses a private format of \textit{Subversion} and neither
\textit{git-svn}\footnote{\url{https://git-scm.com/docs/git-svn}}
nor \textit{svn2git}\footnote{\url{https://github.com/nirvdrum/svn2git}}
can convert this format.
As a result, our experiments apply Bugspots on the \textit{Math},
\textit{Lang}, \textit{Time}, and \textit{Closure} projects.


\subsubsection{Learning to Rank}\label{sec:impl-l2r}
For the learning to rank model, our experiments associate
each program statement with a vector
$$\|Suspiciousness|(e) = \langle s_{t_1}(e), s_{t_2}(e), ... \rangle$$
where $e$ is a program element, and $s_{t_i}(e)$ is the suspiciousness
score of $e$ reported by technique $t_i$.
The vector values are normalized to be within the domain [0, 1], where 1 is
most suspicious and 0 is least suspicious.

Then our experiments apply \emph{rankSVM}~\cite{kuo2014large} to train
the learning to
rank model. \emph{RankSVM} is an open-source learning to rank tool based on
\emph{LIBSVM}~\cite{CC01a}. It implements a pairwise learning to rank
model and has been used in previous fault localization
work~\cite{b2016learning,sohn2017fluccs}. It generates pairwise
constraints, e.g., $e^{\|faulty|}_i > e^{\|correct|}_j$, and the training goal
is to rank the faulty elements at the top, i.e., maximize satisfied
pairwise constraints.

\subsection{Measurements}
To evaluate fault localization techniques, we need
to measure their performance quantitatively.
Previous studies use similar metrics for this measurement, but they
may differ in how they handle cases such as insertion or
multiple faulty elements. This section describes the
measurement methods used in our study.

\subsubsection{Determining Faulty Elements}

To understand how faulty elements are ranked, we need first to
determine which elements in the program are faulty. Following common
practice~\cite{sohn2017fluccs,b2016learning,pearson2017evaluating}, we
define the faulty program elements as those modified or deleted in the
developer patch\footnote{In the Defects4J dataset, the developer patch has
  been minimized to eliminate changes unrelated to the bug fix.}
that fixes the defect. In the
following example patch, i.e., the
\textit{diff} between the fixed and the faulty program, the second line
is considered faulty.


\begin{lstlisting}[moredelim={[is][\bfseries]{@@}{@@}},
    numberstyle=\small\ttfamily,
    basicstyle=\small\ttfamily]
1  if (real == 0.0 && imaginary == 0.0) {
@@2-     return NaN;@@
@@3+     return INF;@@
4  }
5
6  if (isInfinite) {
7      return ZERO;
8  }
9  ...
\end{lstlisting}

However, sometimes a developer patch only inserts new elements rather
than modifies or deletes old elements. To deal with insertions, we
follow the principle used by Pearson et
al.~\cite{pearson2017evaluating}: {a fault localization technique should
  report the element immediately
following the inserted element.}
The rationale is that 
the immediately following element indicates the location that a developer
should change to fix the defect. 

\subsubsection{Multiple Faulty Elements}\label{sec:multi-fault}
Many defective programs have multiple defective elements.

In Defects4J, there are two common reasons for multiple
faulty elements.
\begin{itemize}
\item To repair a fault, the programmer changed multiple
  elements.
\item The patch of a fault not only repairs the current fault but
  also repairs cloned bugs, i.e., the same bug in cloned code snippets. 
\end{itemize}

Following existing work~\cite{pearson2017evaluating}, we consider a
fault to be localized by a fault localization technique if any faulty
element is localized.
It is assumed that if a fault localization technique gives any of the
  faulty elements, the developer can deduce the other faulty
  elements.
Furthermore, when multiple cloned bugs exist, the developer can
re-run fault localization to find the others or can use techniques for repairing
cloned
  bugs~\cite{DBLP:conf/pldi/MengKM11,DBLP:conf/icse/JacobellisMK04}
  to discover and fix other bugs.

\subsubsection{Elements with the Same Score}\label{sec:e_inspect}
Fault localization techniques often
assign the same suspiciousness score to elements, either because the
techniques are designed to only locate elements but not to rank them (e.g.,
union strategy in slicing), or because the techniques cannot distinguish some
elements (e.g., statements in a basic block in SBFL). When presenting the
suspicious list to the user, the elements with the same score are
presented in an arbitrary
order, and thus we need to consider the order when measuring
the performance.

Previous studies~\cite{pearson2017evaluating, wong2016survey,
  steimann2013threats} treat elements with the same score as all the
$n$th element in the list, where $n$ is their \emph{average rank}. However,
this method may unnecessarily lower their ranks when multiple faulty
elements exist. For example, suppose a set of tied elements are
all faulty. Then regardless how this set is sorted, the user will find
a faulty element at the first element in the set, rather than at the
average rank.   

To overcome this problem, in our study we measure the performance of
a fault localization by the expected rank of the first faulty
element, assuming tied elements are arbitrarily sorted.
More concretely, assuming a group of $t$ tied elements
starting at $P_{start}$ that contains $t_{f}$ faulty elements and
there is no faulty element before $P_{\|start|}$, we define
${E_{\|inspect|}}$,
which measures the expected rank of the
first faulty element using the following formula.


\[
E_{\|inspect|} = P_{start} + \sum^{t-t_f}_{k=1} k\frac{\binom {t-k-1}{t_f-1}} {\binom {t}{t_f}}
\]

The equation within the summation is the probability for the top-ranked faulty element
to appear in the $k$th location after $P_{\|start|}$:  that is, the number of all combinations
where the first faulty element is at $k$ ($\binom
  {t-k-1}{t_f-1}$) divided by the number of all combinations ($\binom {t}{t_f}$).

Notice that, when there is only one faulty element, i.e., $t_f = 1$, the equation reduces to:
\[
E_{\|inspect|} = P_{start} + \frac{t - 1}{2}
\]
which is the same as \textbf{average rank}, also \textbf{average accuracy},
in existing
studies~\cite{b2016learning,li2017transforming}.

Also, when all tied
elements are faulty, i.e., $t_f = t$, the equation reduces to:
\[
E_{\|inspect|} = P_{start}
\]
which indicates the first element in the tied set.

\subsubsection{Metrics}\label{sec:metrics}
So far we have defined how to calculate the expected rank of the first
faulty element. Based on this definition we use two metrics to measure
the performance of a fault localization technique.

$\mathbf{E_{\|inspect|}@n}$ counts the number of the 357 faults that were successfully localized
within the top $n$ positions of the resultant ranked lists, i.e., the number of faults that $E_{\|inspect|}$ values on these faults are less than or equal to $n$. It is adapted
from metric
$acc@n$~\cite{b2016learning,li2017transforming}. A previous study~\cite{parnin2011automated}
suggested that programmers will only inspect the top few positions in a
ranked list, and $E_{\|inspect|}@n$ reflects this.

\textbf{\textit{EXAM}}~\cite{wong2008crosstab} is the percentage
of elements that have to be inspected until finding a faulty element,
averaged across all 357 faults uniformly.
It is a commonly used metric for fault localization
techniques~\cite{pearson2017evaluating,naish2011model,wong2012effective}. The
\textit{EXAM} score measures the relative position of the faulty element in
the ranked list. Smaller \textit{EXAM} scores are better.

The ${E_{\|inspect|}@n}$ metric is a more meaningful measure of
fault localization quality than the \textit{EXAM} score.
A developer will only examine the first few
reports from a tool (say, 5 or 10) and a program repair tool will only
examine the first 200 or so reports.  Therefore, any reports other than
these are irrelevant and can be disregarded, yet they are most of the
reports and dominate the \textit{EXAM} score. This paper includes the
\textit{EXAM} score to enable comparison with earlier papers.








\section{Experiment Results}\label{sec:exp-result}
Our experiments investigate and answer six research questions. The granularity
of all experiments is statements, except in
\cref{sec:method-level,sec:compare-other-l2r}.

\subsection{RQ1. Effectiveness of standalone techniques}\label{sec:exp-result-individual}

\subsubsection{Procedure}
To evaluate the effectiveness of standalone fault localization
techniques, we invoked each technique on Defects4J and compared their
$E_{\|inspect|}@n$ and \textit{EXAM} scores.
The definitions of $E_{\|inspect|}@n$ and \textit{EXAM} are in
\cref{sec:metrics}.



\begin{table}[tbp]
  \centering
  \caption{The Performance of Standalone Techniques on all 357 faults.
  Boldface indicates the best-performing techniques.
}
  \label{tab:avg-individual}
  \resizebox{0.49\textwidth}{!}{
  \begin{tabular}{m{4.4em}m{4.5em}|llll|c}
    \toprule
    \multirow{2}{*}{Family} & \multirow{2}{*}{Technique} & \multicolumn{4}{c}{$E_{\|inspect|}$} & \multirow{2}{*}{\textit{EXAM}}\\
    & & @1 & @3 & @5 & @10 & \\
    \midrule
    \multirow{2}{*}{SBFL} & Ochiai & 16 (4\%) & 81 (23\%) & \textbf{111 (31\%)} & \textbf{156 (44\%)} & \textbf{0.033}\\
    & DStar & 17 (5\%) & \textbf{84 (24\%)} & \textbf{111 (31\%)} & 155 (43\%) & \textbf{0.033}\\ \midrule
    \multirow{2}{*}{MBFL} & Metallaxis & 23 (6\%) & 78 (22\%) & 103 (29\%) & 129 (36\%) & 0.118\\
    & MUSE & \textbf{24 (7\%)} & 44 (12\%) & 58 (16\%) & 68 (19\%) & 0.304\\ \midrule
    \multirow{3}{*}{slicing} & union & 5 (1\%) & 33 (9\%) & 58 (16\%) & 84 (24\%) & 0.207\\
    & intersection & 5 (1\%) & 35 (10\%) & 55 (15\%) & 71 (20\%) & 0.222\\
    & frequency & 6 (2\%) & 39 (11\%) & 58 (16\%) & 84 (24\%) & 0.208 \\ \midrule
    stack trace & stack trace & 20 (6\%) & 31 (9\%) & 38 (11\%) & 38 (11\%) & 0.311\\ \midrule
    predicate switching & predicate switching & 3 (1\%) & 15 (4\%) & 20 (6\%) & 23 (6\%) & 0.331\\ \midrule
    IR-based & BugLocator & 0 (0\%) & 0 (0\%) & 0 (0\%) & 3 (1\%) & 0.212 \\  \midrule
    history-based & Bugspots & 0 (0\%) & 0 (0\%) & 0 (0\%) & 0 (0\%) & 0.465\\
    \bottomrule
  \end{tabular}
  }
\end{table}

\subsubsection{Results and Findings}
\Cref{tab:avg-individual} shows the $E_{\|inspect|}@n$ and
\textit{EXAM} of each standalone technique.

\smalltitlecolon{Finding 1.1} SBFL is the most effective
standalone fault localization family in our experiments.


Two techniques of SBFL, Ochiai and DStar, are the best and second
best on $E_{\|inspect|}@3,5,10$ and \textit{EXAM}.
The two techniques locate 156 and 155 faults (about
44\% of all faults) in the top 10 reports. SBFL
may underperform at $E_{\|inspect|}@1$ because blocks are the
minimum granularity that SBFL can identify.
In a single execution, the statements in a basic block are all executed or
not executed, so the elements in a basic block always have the same $\langle
e_f, e_p, n_f, n_p \rangle$ values and the same score.

\smalltitlecolon{Finding 1.2} Bugspots and BugLocator are not as effective at
localizing faulty statements as other techniques.

Bugspots did not locate any fault in its top-10 statements
and BugLocator did not locate any fault in its top-5 statements.
A possible reason is that the two techniques works at the file granularity and
all statements in a file will be tied. Even if the faulty statement is in the
  identified file, the statement will be tied with many other statements
  and will not be ranked high.

\begin{table}[tbp]
  \centering
  \caption{The Performance of Techniques on \textit{Crash Faults} (90 out of 357 faults,
  25\%)
  }
  \label{tab:fault-type}
  \resizebox{0.49\textwidth}{!}{
  \begin{tabular}{m{4.4em}m{4.5em}|llll|c}
  \toprule
  \multirow{2}{*}{Family} & \multirow{2}{*}{Technique} & \multicolumn{4}{c}{$E_{\|inspect|}$} & \multirow{2}{*}{\textit{EXAM}}\\
  & & @1 & @3 & @5 & @10 & \\
  \midrule
  \multirow{2}{*}{SBFL}
  & Ochiai    & 4 (4\%)   & 17 (19\%)  & 32 (36\%)  & \textbf{50 (56\%)}  & \textbf{0.028}\\
  & DStar     & 4 (4\%)   & 18 (20\%)  & 33 (37\%)  & \textbf{50 (56\%)}  & 0.029\\ \midrule
  \multirow{2}{*}{MBFL}
  & Metallaxis& 10 (11\%)  & 30 (33\%)  & 35 (39\%)  & 44 (49\%)  & 0.083\\
  & MUSE      & 6 (7\%)   & 13 (14\%)  & 18 (20\%)  & 19 (21\%)  & 0.345\\ \midrule
  \multirow{3}{*}{slicing}
  & union     & 2 (2\%)   & 13 (14\%)  & 26 (29\%)  & 36 (40\%)  & 0.112\\
  & intersection& 2 (2\%) & 13 (14\%)  & 21 (23\%)  & 30 (33\%)  & 0.136\\
  & frequency & 2 (2\%)   & 14 (16\%)  & 25 (28\%)  & 36 (40\%)  & 0.112\\ \midrule
  stack trace
  & stack trace& \textbf{20 (22\%)} & \textbf{31 (34\%)}  & \textbf{38 (42\%)}  & 38 (42\%)  & 0.194\\ \midrule
  predicate switching
  & predicate switching & 1 (1\%) & 5 (6\%) & 8 (9\%) & 9 (10\%) & 0.323\\ \midrule
  IR-based
  & BugLocator& 0 (0\%)   & 0 (0\%)   & 0 (0\%)   & 0 (0\%)   & 0.199\\ \midrule
  history-based
  & Bugspots  & 0 (0\%)   & 0 (0\%)   & 0 (0\%)   & 0 (0\%)   & 0.433\\
  \bottomrule
  \end{tabular}
  }
\end{table}

\smalltitlecolon{Finding 1.3} Stack trace is the most effective
technique on \emph{crash faults}.

Stack trace analysis only works on crash
faults (\cref{sec:exp-setup-stack}).
In the Defects4J dataset,
25\% of the faults (90 out of 357) are crash faults,
including application-defined exceptions, out of memory errors, and stack
overflow errors.
\Cref{tab:fault-type} shows that
stack trace analysis locates 22\% of
crash faults (20 out of 90) at top-1. By contrast, the second best technique
on these crash faults is Metallaxis, which identifies 11\% of
faults (10 out of 90) at top-1.
This finding indicates that the stack trace is
a vital information source for crash faults. It is also consistent with
debugging scenarios in practice: when a program crashes, the developer
often starts by examining the stack trace.


\begin{table}[t]
  \centering
  \caption{The Performance of Techniques on \textit{Predicate-Related Faults} (115
  out of 357 faults, 32\%)}
  \label{tab:if-fault}
  \resizebox{0.49\textwidth}{!}{
  \begin{tabular}{m{4.4em}m{4.5em}|llll|c}
  \toprule
  \multirow{2}{*}{Family} & \multirow{2}{*}{Technique} & \multicolumn{4}{c}{$E_{\|inspect|}$} & \multirow{2}{*}{\textit{EXAM}}\\
  & & @1 & @3 & @5 & @10 & \\
  \midrule
  \multirow{2}{*}{SBFL}
  & Ochiai    & 5 (4\%)   & 20 (17\%)  & 29 (25\%)  & 43 (37\%)  & \textbf{0.027}\\
  & DStar     & 5 (4\%)   & 21 (18\%)  & 30 (26\%)  & 43 (37\%)  & 0.028\\ \midrule
  \multirow{2}{*}{MBFL}
  & Metallaxis& 8 (7\%)   & \textbf{25 (22\%)}  & \textbf{34 (30\%)}  & \textbf{44 (38\%)}  & 0.090\\
  & MUSE      & \textbf{13 (11\%)}  & 24 (21\%)  & 32 (28\%)  & 35 (30\%)  & 0.174\\ \midrule
  \multirow{3}{*}{slicing}
  & union     & 0 (0\%)   & 6 (5\%)   & 12 (10\%)  & 26 (23\%)  & 0.171\\
  & intersection& 0 (0\%) & 9 (8\%)   & 13 (11\%)  & 20 (17\%)  & 0.185\\
  & frequency & 0 (0\%)   & 10 (9\%)  & 15 (13\%)  & 27 (23\%)  & 0.172\\ \midrule
  stack trace
  & stack trace& 4 (3\%)  & 7 (6\%)   & 10 (9\%)  & 10 (9\%)  & 0.255\\ \midrule
  predicate switching
  & predicate switching & 3 (3\%) & 15 (13\%) & 20 (17\%) & 23 (20\%) & 0.216\\ \midrule
  IR-based
  & BugLocator& 0 (0\%)   & 0 (0\%)   & 0 (0\%)   & 1 (0\%)   & 0.156\\ \midrule
  history-based
  & Bugspots  & 0 (0\%)   & 0 (0\%)   & 0 (0\%)   & 0 (0\%)   & 0.417\\
  \bottomrule
  \end{tabular}}
\end{table}




\smalltitlecolon{Finding 1.4} Predicate switching is not the most effective
technique on ``predicate-related'' faults.

A predicate-related fault is one whose patch modifies the predicate in a
conditional statement. In the Defects4J dataset, 32\% of the faults are
predicate-related faults. 
\Cref{tab:if-fault} shows the performance of each standalone technique on
predicate-related faults. It was surprising to us that MBFL family works
better than predicate switching on predicate-related faults. When working on
predicates, MBFL and predicate switching have similar mechanisms.
They both modify the predicates and check whether the execution result changes.
Predicate switching may underperform MBFL because MBFL can further rank
the critical predicates (modifying which can change the execution result)
while predicate switching cannot.

\subsection{RQ2. Correlation between Techniques}
\label{sec:exp-result-correlation}

This research question explores the possibility of combining different
techniques. Two techniques are (positively) \textbf{correlated} if they are
good at localizing the same sorts of faults. If two
techniques are less correlated, they may provide different information, and
combining them has the potential to outperform either of the component
techniques.


\subsubsection{Procedure}




First, to visually illustrate the correlation between techniques,
we drew the results of each pair of techniques as a scatter plot.
Each figure has 357 points, one for each fault in our dataset. The
coordinate $(x, y)$ for a fault means on this fault, the
$E_{\|inspect|}$
for technique on \textit{X}-axis is $x$, and the
$E_{\|inspect|}$ for technique on \textit{Y}-axis is $y$.



To quantify the correlation between each pair
of techniques, we computed \textit{coefficient of
determination} $r^2$, which
measures of the linear correlation between two
variables~\cite{lee1988thirteen}.
Recall that a developer will only examine the first few
reports from a tool (say, 5 or 10) and a program repair tool will only
examine the first 200 or so reports.
When computing the correlation, we used all points such
that $x \le q$ or $y \le q$, for threshold $q=100$.
We also computed the $p$-value, to determine whether the
correlation coefficient is statistically significant.



\subsubsection{Qualitative Results}\label{sec:overlap-qualitative}


\begin{figure}[tbp]
  \begin{center}
    \includegraphics[scale=0.35]{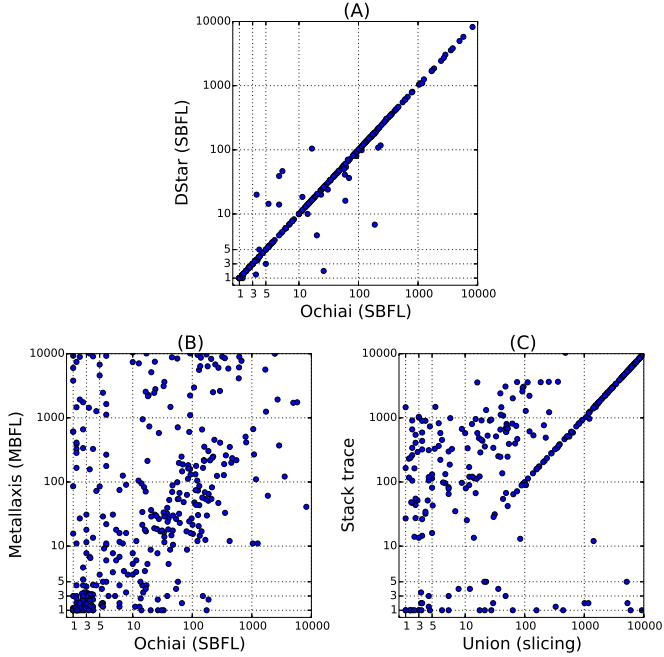}
    \caption{The Correlation of Three Example Pairs of Techniques. The X and Y 
values for a point show the
$E_{\|inspect|}$ values for two techniques on the same bug.
$E_{\|inspect|}$ is the expected rank of the
first faulty element in the FL tool's output, or the number of elements
that a user would have to inspect before inspecting a faulty element.
}

    \label{fig:overlap-all}
  \end{center}
\end{figure}


\Cref{fig:overlap-all}'s scatter plots visualize the correlation between
three sample pairs of techniques. The three pairs capture typical patterns
of the plots, and we omit the rest of the plots as they are similar to
one of the three plots.

\smalltitlecolon{Finding 2.1} Different correlation patterns exist
between different pairs of techniques.

In \cref{fig:overlap-all}~(A), most points lie on the diagonal. This
distribution pattern means the two SBFL techniques, Ochiai and DStar, have
almost the same $E_{\|inspect|}$ values on all faults.
The
two techniques are very correlated and each is unlikely to provide more
information than the other.

In \cref{fig:overlap-all}~(B), there are many faults located in upper-left
and lower-right regions, which correspond to faults that one technique works well
on, while the other works poorly. These faults suggest that the two
techniques are not positively correlated.

The dots in \cref{fig:overlap-all}~(C)
are located on the diagonal in the upper
right region, but they are scattered in other regions. This pattern indicates
that there are a set of faults where both techniques perform poorly,
but there are also many faults where one technique performs well but
the other does not.



\subsubsection{Quantitative Results}\label{sec:overlap-quantitative}

\Cref{tab:similary-r2} shows the \textit{coefficient of determination}, $r^2$,
between each pair of techniques. Different from \cref{fig:overlap-all}, which is
a log-scale plot, this experiment calculates $r^2$ based on $E_{\|inspect|}$,
without log-scale normalization. Notice that the table is a symmetric
matrix.

\smalltitlecolon{Finding 2.2} 
Most techniques are weakly correlated, including all techniques in different families.

In \cref{tab:similary-r2}, there are 55 pairs of different techniques.
Only two of them are significantly correlated at $p$-value less than 0.05
level: $\langle$Ochiai, DStar$\rangle$ from SBFL, with $r^2=0.753$,
$p$-value $\ll 0.01$, and $\langle$\textit{union}, \textit{frequency}$\rangle$ 
from slicing, with $r^2=0.310$, $p$-value $\ll 0.01$.
The $r^2$ values in other pairs of techniques are much smaller, and the
$p$-values of them are larger than 0.05, which suggests that there
is no statistically significant correlation between other pairs
of techniques, at least for the reports that a programmer or tool may view.

Two techniques may provide different information when they are less
correlated. Since there exist many weakly correlated pairs, if a method
could utilize the information from different techniques, it may improve the
effectiveness of fault localization.


\smalltitlecolon{Finding 2.3}
The strongly correlated techniques only exist in the same family, but not all techniques in the same family are strongly correlated.

The most correlated pair of techniques, Ochiai and
DStar, is from the SBFL family. The second most correlated pair is from the
slicing family. However, not all techniques from the same family are
strongly correlated. For example, the two from MBFL family are weakly
correlated, and so is intersection with other slicing techniques. This
finding suggests that though it may be less promising to combine techniques
from the same family, it is still worth investigating.

\begin{table*}[tbp]
  \small
  \centering
  \caption{$r^2$ between Pairs of Techniques. The two pairs of statistically significantly correlated techniques ($p$-value $< 0.05$) are highlighted.}
  \label{tab:similary-r2}
  \resizebox{\textwidth}{!}{
  \begin{tabular}{m{5em}|m{5em}||m{3em}m{3em}|m{4em}m{4em}|m{3em}m{4.2em}m{4.2em}|m{5em}|m{5em}|m{5em}|m{5em}}
    \toprule
    & Family & \multicolumn{2}{c|}{SBFL} & \multicolumn{2}{c|}{MBFL} & \multicolumn{3}{c|}{slicing} & stack trace & predicate switching & IR-based & history-based \\ \midrule
Family & Technique & Ochiai & DStar & Metallaxis & MUSE & union & intersection & frequency & stack trace & predicate switching & BugLocator & Bugspots \\ \midrule\midrule
\multirow{2}{*}{SBFL} & Ochiai & - & \textbf{0.753} & 0.001 & 0.005 & 0.000 & 0.001 & 0.001 & 0.000 & 0.001 & 0.001 & 0.000 \\
                      & DStar  & \textbf{0.753} & - & 0.001 & 0.004 & 0.000 & 0.000 & 0.000 & 0.000 & 0.001 & 0.001 & 0.000 \\ \midrule
\multirow{2}{*}{MBFL} & Metallaxis & 0.001 & 0.001 & 1.000 & 0.002 & 0.008 & 0.005 & 0.005 & 0.004 & 0.003 & 0.003 & 0.001 \\ 
                      & MUSE   & 0.005 & 0.004 & 0.002 & 1.000 & 0.012 & 0.013 & 0.013 & 0.015 & 0.009 & 0.015 & 0.024 \\ \midrule
\multirow{3}{*}{slicing} & union & 0.000 & 0.000 & 0.008 & 0.012 & - & 0.004 & \textbf{0.310} & 0.009 & 0.010 & 0.003 & 0.000 \\ 
                      & intersection & 0.001 & 0.000 & 0.005 & 0.013 & 0.004 & - & 0.015 & 0.007 & 0.009 & 0.004 & 0.003 \\ 
                      & frequency & 0.001 & 0.000 & 0.005 & 0.013 & \textbf{0.310} & 0.015 & - & 0.005 & 0.009 & 0.003 & 0.000 \\ \midrule
stack trace & stack trace & 0.000 & 0.000 & 0.004 & 0.015 & 0.009 & 0.007 & 0.005 & - & 0.014 & 0.005 & 0.022 \\ \midrule
predicate switching & predicate switching & 0.001 & 0.001 & 0.003 & 0.009 & 0.010 & 0.009 & 0.009 & 0.014 & - & 0.007 & 0.026 \\ \midrule
IR-based & BugLocator & 0.001 & 0.001 & 0.003 & 0.015 & 0.003 & 0.004 & 0.003 & 0.005 & 0.007 & - & 0.009 \\ \midrule
history-based & Bugspots & 0.000 & 0.000 & 0.001 & 0.024 & 0.000 & 0.003 & 0.000 & 0.022 & 0.026 & 0.009 & - \\ \midrule
    \bottomrule
  \end{tabular}
  }
\end{table*}

\subsection{RQ3. Effectiveness of Combining Techniques}
\label{sec:exp-result-combined}
\Cref{sec:exp-result-correlation} indicates that the techniques are potentially complementary
to each other.
This section applies the learning to rank model to combine
techniques.

\subsubsection{Procedure}



Our experiments
perform cross-validation to evaluate
the ranking model. 
Cross-validation estimates model performance without losing modeling or test
capability despite small data size. 
In particular, we used two cross-validation methods.
\begin{itemize}
\item {\it $k$-fold validation.} This simulates within-project
  training. The original data were randomly split into $k$
different sets of the same size and the training-validation is
performed $k$ times, each time training on $k$-1 sets and validating on
the other set. We set $k=10$ in our experiment.
\item {\it Cross-project validation.} This simulates cross-project
  training. We treat one project as the test set and 
the other projects as the
training sets, and repeat the process for each project.
\end{itemize}





We performed two sets of experiments to evaluate the combined technique.
\begin{itemize}
\item The first experiment measured the performance of combining all techniques. 
\item The second experiment evaluated the contribution of each
  fault localization family. We excluded one family at a time and repeated the learning
  to rank procedure.
\end{itemize}

\begin{table}[tbp]
  \centering
  \caption{Learning to Rank Results. Comparing cross project validation and $k$-fold validation.}
  \label{tab:l2r-sanitycheck}
  \resizebox{0.48\textwidth}{!}{
  \begin{tabular}{l|llll|l}
    \toprule
    \multirow{2}{*}{Validation Method} & \multicolumn{4}{c}{$E_{\|inspect|}$} & \multirow{2}{*}{\textit{EXAM}}\\
    & @1 & @3 & @5 & @10\\
    \midrule
    10-fold   & 72 (20\%) & 137 (38\%) & 168 (47\%) & 205 (57\%) & 0.0173\\
    cross project & 68 (19\%) & 130 (36\%) & 165 (46\%) & 197 (55\%) & 0.0171\\
    \bottomrule
  \end{tabular}
  }
\end{table}

\newcommand{\without}{w$\!$/$\!$o\xspace}

\begin{table}[tbp]
  \centering
  \caption{
  Learning to Rank Results. Learning to rank is significantly
    better than any original technique. The reduction of excluding a
    family is marked after the $E_{\|inspect|}$ value.
  The Ochiai, DStar, and MUSE rows are copied from \cref{tab:avg-individual} for comparison.}
  \label{tab:l2r-result}
  \resizebox{0.48\textwidth}{!}{
  \setlength{\tabcolsep}{.8\tabcolsep}
  \begin{tabular}{m{8em}|llll|l}
    \toprule
    Family / & \multicolumn{4}{c}{$E_{\|inspect|}$} & \multirow{2}{*}{\textit{EXAM}}\\
    Technique & @1 & @3 & @5 & @10\\
    \midrule
    All Families          & 72 (20\%) & 137 (38\%) & 168 (47\%) & 205 (57\%) & 0.0173\\
    \midrule
    \without SBFL         & 61 (-11) & 120 (-17) & 145 (-23) & 188 (-27) & 0.0225 \\
    \without MBFL         & 52 (-20) & 122 (-15) & 148 (-20) & 194 (-11) & 0.0206 \\
    \without slicing      & 58 (-14) & 129 (-8) & 165 (-3) & 201 (-4) & 0.0190 \\
    \without stack trace  & 63 (-9) & 133 (-4) & 161 (-7) & 199 (-6) & 0.0176 \\
    \mbox{\without predicate} \mbox{\quad switching} & 68 (-4) & 136 (-1) & 165 (-3) & 198 (-7) & 0.0178 \\
    \without IR-based     & 66 (-6) & 134 (-3) & 162 (-6) & 194 (-11) & 0.0173 \\
    \without history-based & 71 (-1) & 136 (-1) & 167 (-1) & 203 (-2) & 0.0173 \\
    \midrule\midrule
    Ochiai & 16 (4\%) & 81(23\%) & {111 (31\%)} & {156 (44\%)} & {0.033}\\
    DStar & 17 (5\%) & {84 (24\%)} & {111 (31\%)} & 155 (43\%) & {0.033}\\
    MUSE & {24 (7\%)} & 44 (12\%) & 58 (16\%) & 68 (19\%) & 0.304\\
    \bottomrule
  \end{tabular}
  }
\end{table}

\subsubsection{Results and Findings}

\smalltitlecolon{Finding 3.1} The two cross-validation methods yield similar evaluation
results.

\Cref{tab:l2r-sanitycheck} shows the results of the first experiment on
validation methods. The evaluation results of the two validation methods are
similar. Both of the two validation methods indicate that the combined technique
significantly outperforms any standalone techniques in
\Cref{tab:avg-individual}. 
These results suggest that the learning to rank model we used has
good generalizability across different projects. Since the performances
of the two methods are close, the rest of the paper reports only
10-fold validation. 

\smalltitlecolon{Finding 3.2} The combined technique
significantly outperforms any standalone technique.

\Cref{tab:l2r-result} shows the results of the two experiments on combined
techniques. The
\textit{All Families} row presents the results of the first experiment, i.e.,
the results of combining all families. The next rows present
the results of the second experiment, where each row shows the performance of
excluding one family at a time. The \textit{reduction} of excluding a family is
marked after the $E_{\|inspect|}$ value.


The combined technique in the \textit{All families} row is
significantly better than any standalone techniques. At
$E_{\|inspect|}@1,3,5,10$, the combined technique improves 200\%, 63\%,
46\% and 31\% over the former best, respectively. At \textit{EXAM}, it
improves from 0.033 to 0.0173, an improvement of 48\% from the former
best. These results indicate that learning to rank is an effective method
to combine different fault localization techniques and the performance of
the combined technique is significantly improved.


\smalltitlecolon{Finding 3.3}
The contribution of each technique to the combined result is
not determined by its effectiveness as a standalone technique.

For example, while the IR-based family could not locate any bugs in Top
1--5 and predicate switching can locate 3--20 bugs in Top 1--5, removing
the IR-based family has a larger impact than predicate switching in Top
1--5. This finding indicates that, when considering a fault localization
technique, it is not enough to evaluate its individual performance: we need
to evaluate it in combination with other techniques.




\smalltitlecolon{Finding 3.4} All families contribute to the overall
results.

\Cref{tab:l2r-result} shows that removing any family 
decreases all metrics.
Bugspots, which does not rank any faulty element into the top 10 when used
alone, slightly improved all $E_{inspect}@n$ values when
combined with other techniques.

\begin{table*}[tbp]
  \centering
  \caption{Time Consumption of Each Technique (in seconds, to 2 digits of precision)}
  \label{tab:time-consume}
  \begin{tabular}{c|m{6em}|m{6em}|r|rrrrr}
  \toprule
  Time Level & Family & Technique & Average & Math & Lang & Time & Chart & Closure \\
  \midrule
  \multirow{3}{*}{Level 1 (Seconds)} & history-based & Bugspots 
                    & 0.54 & 0.66 & 0.22 & 0.20 & - & 0.67 \\ \cmidrule{2-9}
  & stack trace & stack trace 
                    & 1.3 & 0.17 & 0.15 & 0.39 & 0.18 & 3.1 \\ \cmidrule{2-9}
  & IR-based & BugLocator
                    & 5.6 & 6.6 & 4.3 & 4.7 & 4.6 & 5.8 \\ \midrule
  \multirow{5}{*}{Level 2 (Minutes)} & \multirow{3}{*}{slicing} 
                    & union & 80 & 44 & 39 & 29 & 47 & 150 \\
  & & intersection  & 80 & 44 & 39 & 29 & 47 & 150 \\
  & & frequency     & 80 & 44 & 39 & 29 & 47 & 150 \\ \cmidrule{2-9}
  & \multirow{2}{*}{SBFL} & Ochiai 
                    & 200 & 86 & 26 & 85 & 44 & 430 \\ 
  & & DStar         & 200 & 86 & 26 & 85 & 44 & 430 \\ \midrule
  Level 3 (Around ten minutes) & predicate switching & predicate switching 
                    & 620 & 170 & 73 & 1100 & 120 & 1200 \\ \midrule 
  \multirow{2}{*}{Level 4 (Hours)} & \multirow{2}{*}{MBFL} & Metallaxis 
                    & 4800 & 3000 & 270 & 12000 & 5400 & 7000 \\ 
  &                 & MUSE & 4800 & 3000 & 270 & 12000 & 5400 & 7000 \\ 
  \midrule
  \midrule
  - & \multicolumn{2}{c|}{learning to rank} & 11 & 0.32 & 0.082 & 0.68 & 0.42 & 28 \\ \bottomrule
  \end{tabular}
\end{table*}


\subsection{RQ4. Time Consumption and Combination Strategy}
\label{sec:exp-result-time}

This research question measures the run-time cost of each technique.
Furthermore, we explored the optimal combination strategy under different
time limitations, corresponding to various debugging scenarios.

\subsubsection{Procedure}

We designed two experiments. The first experiment measured
the time consumption for each fault localization technique. We also
measured the run time for the learning to rank model, which
indicates the combination overhead. The second experiment combined fault
localization families one by one and 
measured the execution time and the performance of the combined
technique in order to find optimal combinations under different time limits.

Our experiments include or exclude an entire family at a time, rather than
including/excluding specific techniques.  The reason is that for each
family, all techniques use the same raw data.  Once the raw data is
collected, the overhead for applying an extra technique from the same
family is only re-calculating the scores and re-ranking the program
elements, which is negligible.


\subsubsection{Results and Findings}

\Cref{tab:time-consume} shows the time consumption for each
technique. The \textit{average} column presents the
average time consumed per fault over
the whole dataset, and the \textit{project name} columns present the run time
for the specific project.

\smalltitlecolon{Finding 4.1} The training time for learning to rank
  is small compared to the fault localization time.

The \textit{learning to rank} row at the bottom of \cref{tab:time-consume} shows the
overhead for the training procedure, which costs around 10 seconds on
average. The combination time is always less than a second except that it
is 28 seconds for Closure. A possible reason is Closure is a JavaScript compiler
and FL techniques would generate a long suspicious list, which make the learning
procedure takes longer run-time.
Since the combination of techniques involves at least two
different techniques, this result suggests the overhead introduced by
learning to rank model is small.



\smalltitlecolon{Finding 4.2} The efficiency of families can be categorized into several levels with different orders of magnitude.


\begin{itemize}
  \item Level 1: history-based, stack trace, and IR-based. 
  Bugspots is the fastest technique; it only needs to examine the 
  development history. Stack trace is also a fast technique;
  it needs to \textbf{execute} the test cases, \textbf{once}.
  IR-based technique measures the textual similarity
  between the bug report and the source files, which takes
  a few seconds.

  \item Level 2: slicing and SBFL\@. The slicing and SBFL families
  have similar mechanisms. They need to \textbf{trace} the
  execution of test cases, \textbf{once}. The main difference
  that affects the efficiency is that SBFL needs to trace all the
  test cases while slicing only needs to trace failed test
  cases.

  \item Level 3: predicate switching. Predicate switching is
  slower than the above families; it needs to \textbf{modify
  predicates} in the program and execute test cases
  \textbf{multiple times}.

  \item Level 4: MBFL\@. MBFL is the slowest family;
  it needs to \textbf{modify all possible statements} in the
  program and execute test cases \textbf{multiple times}.
\end{itemize}

\begin{table*}[tbp]
  \centering
  \caption{Optimal Strategies under Time Consumption Levels}
  \label{tab:time-level}
  \begin{tabular}{l|l|c|llll|l}
  \toprule
  \multirow{2}{*}{Time Level} & \multirow{2}{*}{Technique} & Estimated Time & \multicolumn{4}{c}{$E_{\|inspect|}$} & \multirow{2}{*}{\textit{EXAM}}\\
    & & (in seconds) & @1 & @3 & @5 & @10\\
  \midrule
  \multirow{4}{*}{Level 1} 
          & history-based & 0.54 & 0 (0\%) & 0 (0\%) & 0 (0\%) & 0 (0\%) & 0.465 \\
          & stack trace & 1.3 & 19 (5\%) & 29 (8\%) & 35 (10\%) & 35 (10\%) & 0.311 \\
          & stack trace +history-based & 13 & 19 (5\%) & 29 (8\%) & 35 (10\%) & 35 (10\%) & 0.311 \\
          & \textbf{stack trace +history-based +IR-based} & 19 & \textbf{25 (7\%)} & \textbf{42 (12\%)} & \textbf{53 (15\%)} & \textbf{63 (18\%)} & \textbf{0.0421} \\
  \midrule
  \multirow{3}{*}{Level 2}
  & Level 1 +slicing       & 98  & 28 (8\%) & 65 (18\%) & 95 (27\%) & 124 (35\%) & 0.0353 \\
  & Level 1 +SBFL          & 220 & 39 (11\%) & 105 (29\%) & 132 (37\%) & 174 (49\%) & 0.0244 \\
  & \textbf{Level 1 +SBFL +slicing} & 300 & \textbf{52 (15\%)} & \textbf{120 (34\%)} & \textbf{146 (41\%)} & \textbf{189 (53\%)} & \textbf{0.0217} \\
  \midrule
  Level 3
  & Level 2 +predicate switching & 920 & 52 (15\%) & 122 (34\%) & 148 (41\%) & 194 (54\%) & 0.0206 \\
  \midrule
  Level 4 & Level 3 +MBFL & 5700 & 72 (20\%) & 137 (38\%) & 168 (47\%) & 205 (57\%) & 0.0173 \\
  \bottomrule
  \end{tabular}
\end{table*}

\smalltitlecolon{Finding 4.3} Including preceding
level families only slightly affects the time consumption but always
improves the results.  Therefore, all techniques in preceding levels should
be included.

\Cref{tab:time-level}
shows the combinations at different time consumption levels, and the estimated
time consumption. If more than one family is included, the estimated time
consumption is the running time for each family and the training time for
learning to rank. For each level, we merged the corresponding families
into the preceding time levels one by one. 

\Cref{tab:time-level} shows that performance is
significantly improved from level 1 to level 2. This result means
slicing and SBFL brings vital information to the combined technique.
It is also
notably improved from level 3 to level 4, which means MBFL brings useful information to the
combined technique, but it is also very costly.

Using \cref{tab:time-level}, developers can pick the best
combination of techniques based on their use case.
If the fault is a crash fault,
the developer may try level 1 first, which gives the result instantly
and is effective for crash bugs.
For other real-time debugging, developers should try the combination at
level 2, which only takes a few minutes.
Since level 3 is three times
as expensive as level 2 but the results are barely different, a
developer would never choose to run level 3.
If a developer debugs for more than a few minutes, it makes sense to run
level 4 in the background and examine its results as soon as they are
available, since CPU costs are much lower than human time.




\smalltitlecolon{Finding 4.4} Level 2 and Level 4 are two levels with
good balance between effectiveness and efficiency, while Level 1 is
a good choice for crash bugs.


\subsection{RQ5. Results at Method Granularity}\label{sec:method-level}

\Cref{sec:exp-result-individual,sec:exp-result-correlation,sec:exp-result-combined,sec:exp-result-time}
answered the RQs at statement granularity.
Some other
studies have suggested that method may be a better granularity for
developers~\cite{b2016learning,poshyvanyk2007feature}.
We repeated the previous experiments at method granularity and checked
whether the answers still hold.

The suspiciousness score for a method is defined as the maximum score of
its statements.

\subsubsection{Results and Findings}

\smalltitlecolon{Finding 5.1} The main findings in RQ1 and RQ3 still hold at method granularity.

\Cref{tab:avg-individual-method} shows the $E_{\|inspect|}@n$ and
\textit{EXAM} for each technique. The \textit{EXAM} here presents the
percentage of methods needed to inspect before finding the faulty one. The
findings in RQ1 still hold at method granularity:
\begin{itemize}
  \item SBFL is the most effective fault localization family. Ochiai and DStar
  have the best performance on all metrics.
  \item Stack trace is the most effective technique on \textit{crash faults}. Based on 88
  crash faults, stack trace can locate 44\% of them at top-1, and 83\% at
  top-10, which is consistent with the previous study~\cite{schroter2010stack}.
  \item The relative performance between techniques have no significant changes.
\end{itemize}

\Cref{tab:l2r-result-method} shows the results of the learning to
rank model. The results are significantly improved from standalone
techniques in \cref{tab:avg-individual-method}, which is consistent with
the main findings in RQ3.




\begin{table}[tbp]
  \centering
  \caption{The Performance of Standalone Techniques, Method Granularity.
  }
  \label{tab:avg-individual-method}
  \resizebox{0.49\textwidth}{!}{
  \begin{tabular}{m{4.4em}m{4.5em}|llll|c}
    \toprule
    \multirow{2}{*}{Family} & \multirow{2}{*}{Technique} & \multicolumn{4}{c}{$E_{\|inspect|}$} & \multirow{2}{*}{\textit{EXAM}}\\
    & & @1 & @3 & @5 & @10 & \\
    \midrule
    \multirow{2}{*}{SBFL}
    & Ochiai & 92 (26\%) & 180 (50\%) & \textbf{207 (58\%)} & \textbf{241 (68\%)} & \textbf{0.044}\\
    & DStar & \textbf{95 (27\%)} & \textbf{182 (51\%)} & \textbf{207 (58\%)} & \textbf{241 (68\%)} & \textbf{0.044}\\ \midrule
    \multirow{2}{*}{MBFL}
    & Metallaxis & 83 (23\%) & 151 (42\%) & 181 (51\%) & 208 (58\%) & 0.108\\
    & MUSE & 54 (15\%) & 95 (27\%) & 112 (31\%) & 134 (38\%) & 0.274\\ \midrule
    \multirow{3}{*}{slicing}
    & union & 35 (10\%) & 80 (22\%) & 106 (30\%) & 131 (37\%) & 0.259\\
    & intersection & 35 (10\%) & 73 (20\%) & 90 (25\%) & 114 (32\%) & 0.279\\
    & frequency & 39 (11\%) & 84 (24\%) & 104 (29\%) & 133 (37\%) & 0.259\\ \midrule
    stack trace 
    & stack trace & 39 (11\%) & 59 (17\%) & 68 (19\%) & 73 (20\%) & 0.366\\ \midrule
    predicate switching 
    & predicate switching & 15 (4\%) & 38 (11\%) & 50 (14\%) & 60 (17\%) & 0.390\\ \midrule
    IR-based
    & BugLocator & 0 (0\%) & 3(1\%) & 11(3\%) & 35(10\%) & 0.275\\ \midrule
    history-based
    & Bugspots & 0 (0\%) & 2 (1\%) & 4 (1\%) & 13 (4\%) & 0.498\\
    \bottomrule
    \multicolumn{6}{l}{$^*$ \textit{EXAM} here is based on the number of methods.}
  \end{tabular}
  }
\end{table}


\begin{table}[tbp]
  \centering
  \caption{Learning to Rank Results, Method Granularity.
  }
  \resizebox{0.48\textwidth}{!}{
  \label{tab:l2r-result-method}
  \begin{tabular}{p{55pt}|llll|c}
    \toprule
    \multirow{2}{*}{Technique} & \multicolumn{4}{c}{$E_{\|inspect|}$} & \multirow{2}{*}{\textit{EXAM}}\\
     & @1 & @3 & @5 & @10\\
    \midrule
    All techniques & 168 (47\%) & 230 (64\%) & 247 (69\%) & 271 (76\%) & 0.0034 \\
    \bottomrule
  \end{tabular}}
\end{table}

\subsection{RQ6. Comparison with State-of-the-Art Techniques}\label{sec:compare-other-l2r}
Other recent learning to rank
approaches~\cite{xuan2014learning,b2016learning,sohn2017fluccs,li2017transforming}
improve the performance of fault localization by combining techniques in
one family or by augmenting one family with additional information.
We compared our approach with these techniques. A detailed discussion of
the compared techniques can be found in \cref{sec:related-l2r}.

We obtained the performance of the compared approaches on Defects4J from previous publications~\cite{b2016learning,sohn2017fluccs,li2017transforming}. Three of them (MULTRIC, Savant, TraPT) were evaluated on the whole dataset of Defects4J, while FLUCCS was evaluated on a subset of Defects4J containing 210 faults. To compare with FLUCCS, we also performed a cross-validation of our approach over the subset of 210 faults. All results of the compared approaches were obtained via cross-validation, where FLUCCS uses 10-fold cross-validation, and MULTRIC, Savant, and TraPT use 357-fold cross-validation.

This paper used newly defined metrics $E_{\|inspect|}$ at top-n and others
used \textit{average rank} at top-n.  These two metrics are only equivalent
when $n=1$, so \cref{tab:l2r-other-l2r} shows that.
All results are at the method granularity as all the compared approaches support only method granularity.

The result in~\cref{tab:l2r-other-l2r} shows that CombineFL, which is the approach proposed in this paper, is significantly better than all these techniques. This result indicates that 
combining techniques from different families is an effective way to improve the performance of fault localization approaches. Furthermore, some information used in the compared approaches are not used in our approach, so we may further combine these techniques to achieve potentially better results in the future.


Notice that the aforesaid discussion only compares the \textit{output}
between the approaches. In practice, the \textit{run-time cost} is also an
important metric when comparing approaches. For example, since FLUCCS does
not include the mutation component, it might require significantly less
execution time than CombineFL\@. However, the existing papers did not report
the run-time cost of these approaches so a comparison is left for future work.

\begin{table}[tbp]
  \centering
  \caption{Comparison with other learning to rank techniques, for
    $E_{\|inspect|}@1$ at method granularity.
    CombineFL is the approach proposed in this paper.
  }
  \label{tab:l2r-other-l2r}
  \begin{tabular}{l|cc}
  \toprule
  Technique & on 357 faults & on 210 faults \\ \midrule
  CombineFL & \textbf{168} & \textbf{118} \\ \midrule
  MULTRIC~\cite{li2017transforming} & 72 & - \\
  Savant~\cite{b2016learning} & 63 & - \\
  TraPT~\cite{li2017transforming} & 142 & - \\
  FLUCCS~\cite{sohn2017fluccs} & - & 106 \\
  \bottomrule
  \end{tabular}
\end{table}

\section{Related Work}\label{sec:related}

To our knowledge, this paper is the first empirical study on a wide
range of fault localization families.

\subsection{Learning to Combine}\label{sec:related-l2r}

Several studies have applied the learning to rank model to improve the
effectiveness of fault localization techniques.

Xuan and Monperrus~\cite{xuan2014learning} proposed a learning-based
approach, MULTRIC, to integrate 25 existing SBFL risk formulae. They
conducted experiments on ten open-source Java programs with 5386
seeded (artificial) faults, and found that MULTRIC is more effective
than theoretically optimal formulae studied by Xie et
al.~\cite{xie2013theoretical}. In this paper, we found that different
techniques in SBFL family may contain strongly correlated information on
real-world projects. To further improve the fault localization
effectiveness, extra information sources should be introduced rather
than only considering the SBFL family.


Le et al.~\cite{b2016learning} presented Savant, which augmented SBFL
with Daikon~\cite{ernst2001dynamically} invariants as an additional
feature. They applied the learning to rank model to integrate SBFL
techniques and invariant information. They evaluated Savant on
real-world faults from the Defects4J~\cite{just2014defects4j}
dataset and found that Savant outperforms the best four SBFL
formulae, including MULTRIC\@.

Sohn and Yoo~\cite{sohn2017fluccs} proposed FLUCCS, which extended
SBFL techniques with code change metrics. They applied two learning
to rank techniques, Genetic Programming, and linear rank Support
Vector Machines. They also evaluated FLUCCS on the Defects4J dataset
and found FLUCCS exceeds state-of-the-art SBFL
techniques.

Li and Zhang~\cite{li2017transforming} proposed TraPT, which used the
learning to rank technique to extend MBFL with mutation information
gathered from test code and messages. In their experiments,
TraPT outperformed state-of-the-art MBFL and SBFL
techniques.

To sum up, existing studies mainly focus on combining
techniques in one family or augmenting one family with additional
information. Compared with these studies, this paper is the first comprehensive
and systematic 
study to combine a wide range of families. Our
study includes eleven techniques from seven families,
and we analyzed the contribution and the cost of
each technique. The combined technique significantly outperforms any
standalone technique. Nevertheless, we also observe that existing studies
use some information that has not been considered in this paper.
The additional information
could further improve to the combined technique. 





\subsection{Empirical Studies on Fault Localization}
Fault localization techniques have been extensively evaluated
empirically.

Jones and Harrold~\cite{jones2005empirical} introduced the Tarantula
SBFL technique and compared it with three other SBFL techniques based on
test coverage (Set
Union, Set Intersection, Nearest-Neighbor~\cite{renieres2003fault}) and
with Cause Transitions~\cite{cleve2005locating} on the
\textit{Siemens test suite}~\cite{hutchins1994experiments}.
They found that Tarantula is more effective and efficient than the other
techniques.

Abreu et al.~\cite{abreu2007accuracy} introduced Ochiai, another SBFL
technique.  They found that Ochiai outperforms two other SBFL techniques
(Jaccard~\cite{chen2002pinpoint} and
Tarantula~\cite{jones2005empirical})
on the \textit{Siemens test suite}.

Le et al.~\cite{le2013theory} also empirically evaluated several SBFL
techniques on the \textit{Siemens test suite}, to check whether the
theoretically and practically best SBFL techniques match. This study
suggested that Ochiai outperforms the theoretically optimal techniques
by Xie et al.~\cite{xie2013theoretical}, because the optimality
assumptions are unmet on their dataset.

Wong et al.~\cite{wong2014dstar} introduced DStar and compared over
thirty SBFL techniques on nine different sets of programs, including
\textit{Siemens test suite} and several other projects.
They found that DStar is more effective than all other techniques on
all projects.

Pearson et al.~\cite{pearson2017evaluating} evaluated SBFL and MBFL techniques
on both artificial and real-world faults to
find whether the previous findings over artificial faults still hold on 
real-world faults. They identify several cases where results on artificial faults are
different from those on real-world faults, indicating that experimenting over
real-world faults is important.
In other words, results from the Siemens test suite are not characteristic
of real-world faults.

Zhang et al.~\cite{zhang2007study} evaluated three dynamic slicing
techniques on a set of real-world faults. They found that data
slicing~\cite{zhang2003precise} is effective for memory related
faults and full slicing~\cite{korel1988dynamic} was adequate for
other faults. None of the faults in
their dataset required Relevant
slicing~\cite{agrawal1993incremental,gyimothy1999efficient}.

To sum up, existing studies mainly focus on evaluating techniques in
one family, in particular, the SBFL family. Compared with these
studies, our work evaluates a wide range of seven families. In addition,
we also evaluate on large real-world projects and use a new metric to
better measure elements with the same score. Finally, we also
evaluate the combination of different techniques.










\section{Implications}\label{sec:implication}
This section highlights implications for future
research in fault localization.

\subsection{Evaluating Fault Localization Techniques}
Traditionally, fault localization techniques are often used and evaluated
individually.
This paper shows that it is easy to combine even very different fault
localization techniques.  We recommend that users should
not to use a technique standalone, but instead combine multiple
techniques within a time limit.

This implies that, for researchers evaluating a fault localization
technique, it is more important to understand how the technique
contributes in combination with existing techniques, than 
understanding the performance of the technique in isolation.
Understanding the combination includes two aspects: (1)
how much this technique can contribute to the combination of all
existing approaches, and (2) how much this technique can contribute to
the combination within a specific time limit. That is, both
effectiveness and efficiency should be considered.

\subsection{Infrastructure for Evaluating Fault Localization
  Techniques}
To facilitate evaluation of future fault localization techniques,
our infrastructure CombineFL-core and the fault localization data
of the eleven studied fault localization techniques are available
at \url{https://damingz.github.io/combinefl/index.html}.

Given a user-selected
combination of techniques, our infrastructure automatically calculates
its $E_{\|inspect|}$ and \textit{EXAM} scores on Defects4J and measures the
execution time. To integrate a new technique into the dataset, the
user only needs to provide the suspiciousness scores for program elements
in each defect, as well as the execution time, in a specific format.
Then the combinations of the newly added technique with any other
existing techniques are automatically supported. Both statement
granularity and method granularity are supported.

\subsection{Efficiency}
In the existing evaluation of fault localization approaches, efficiency
often receives less attention than effectiveness. However, our study
reveals that different techniques have huge differences in
execution time, and some techniques are
infeasible in certain use cases. Thus, efficiency
is a critical issue that must be taken into consideration when
evaluating fault localization techniques. Furthermore, 
optimizing the efficiency of fault localization
techniques~\cite{wang2017faster,wang2017cost} is an important research
direction.

On the other hand, it is so far not clear how exactly efficiency affects
the debugging performance of developers. This relates to questions such as:
is it worthwhile to wait for the fault localization technique to produce a
more accurate result or should the developer start with a less accurate
result? Future work is needed to answer these questions.

\subsection{Information Sources}
Our study reveals that, when two techniques use the same information
source, their performance is similar. Thus, in 
fault localization research, it seems to be more promising
to find new information sources than optimizing existing
information sources. Recent
studies~\cite{b2016learning,sohn2017fluccs,li2017transforming} also
confirm that integrating more information sources
significantly outperforms any techniques in the SBFL family.

\subsection{Methods for Combining Approaches}
Our study used a learning to rank approach to combine
different techniques. This approach treats different techniques as
black boxes and combines the suspiciousness scores linearly. 
This simple approach has multiple limitations. First,
treating different techniques as black boxes disallows fine-grained
combination. For example, different techniques may contain the same
computations, but treating them as black boxes does not allow us to
reuse these computations nor to
utilize any intermediate results. Second, linear combination may not
be optimal, and other possibilities are left to be explored.
Third, this approach requires a training process, and how much the training
data affect the effectiveness is yet unknown. These limitations call
for new research on novel ways to combine different techniques as well
as understanding more about the learning to rank approach.





\section{Conclusion}\label{sec:conclusion}



This paper investigates the performance of a wide range of fault localization
techniques, including eleven techniques from seven families, on 357 real-world faults.
We evaluated the effectiveness of each standalone fault localization
technique.
Then we applied learning to rank model to combine
these fault localization techniques. Finally, we also measured the execution
time. Our experiments included both statement and method granularities.

The combined techniques  
significantly outperform any standalone technique.
Furthermore, different techniques have significant different
execution time. Based on these findings, we recommend
combining fault localization techniques grouped by different time cost levels, and future fault
localization techniques should also be evaluated in this setting. To
facilitate research and application, 
our infrastructure CombineFL-core and the fault localization data of the
eleven fault localization techniques for evaluating and combining fault
localization techniques 
is available at \url{https://damingz.github.io/combinefl/index.html}.







%



\ifCLASSOPTIONcompsoc
  \section*{Acknowledgments}
\else
  \section*{Acknowledgment}
\fi

This material is based on research sponsored by the
National Key Research and Development Program of China 
No. 2017YFB1001803, the National Natural Science Foundation
of China under Grants Nos. 61529201, 61672045, and the Air Force Research
Laboratory and DARPA of the U.S.A.\
under agreement numbers FA8750-12-2-0107,
FA8750-15-C-0010, and FA8750-16-2-0032. The U.S. Government is authorized
to reproduce and distribute reprints for Governmental purposes
notwithstanding any copyright notation thereon.

\ifCLASSOPTIONcaptionsoff
  \newpage
\fi



\bibliographystyle{IEEEtran}
\bibliography{reference}
%




%
\begin{IEEEbiography}[{\includegraphics[width=1in,height=1.25in,clip,keepaspectratio]{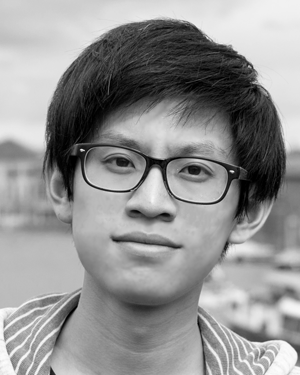}}]{Daming Zou}
received his B.S.\ degree in computer science and technology from Peking 
University. He is currently working toward the Ph.D.\ degree advised by
Professor Lu Zhang and Professor Yingfei Xiong at Peking University. His
research interests include software testing, software analysis, and program repair.
\end{IEEEbiography}



\begin{IEEEbiography}[{\includegraphics[width=1in,height=1.25in,clip,keepaspectratio]{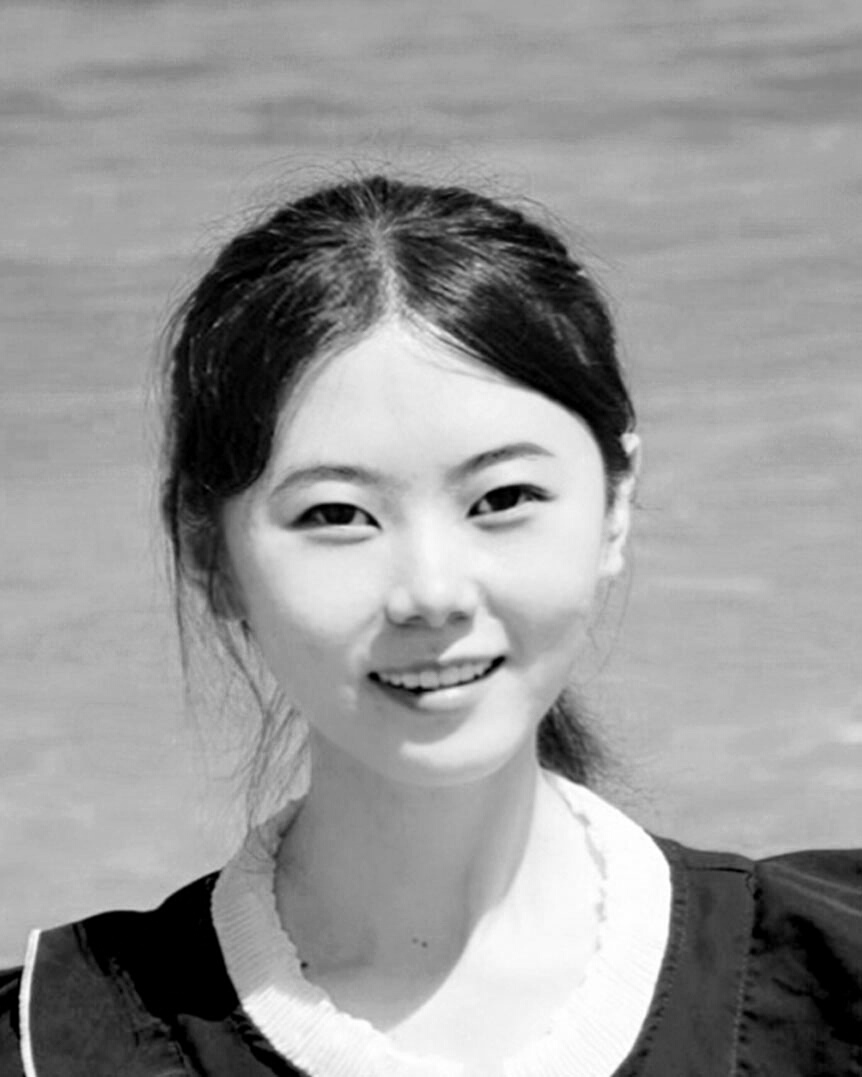}}]{Jingjing Liang}
is a Ph.D.\ Student advised by Professor Yingfei Xiong at
Peking University.
She received her B.S.\ degree in software engineering from East China Normal
University.
Her research interests are fault localization and program repair.
\end{IEEEbiography}


\begin{IEEEbiography}[{\includegraphics[width=1in,height=1.25in,clip,keepaspectratio]{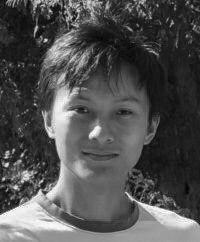}}]{Yingfei Xiong}
is an associate professor at Peking University.
He got his Ph.D.\ degree from the University of Tokyo in 2009 and worked as a
postdoctoral fellow at University of Waterloo between 2009 and 2011. His
research interest is software engineering and programming languages.
\end{IEEEbiography}

\begin{IEEEbiography}[{\includegraphics[width=1in,height=1.25in,clip,keepaspectratio]{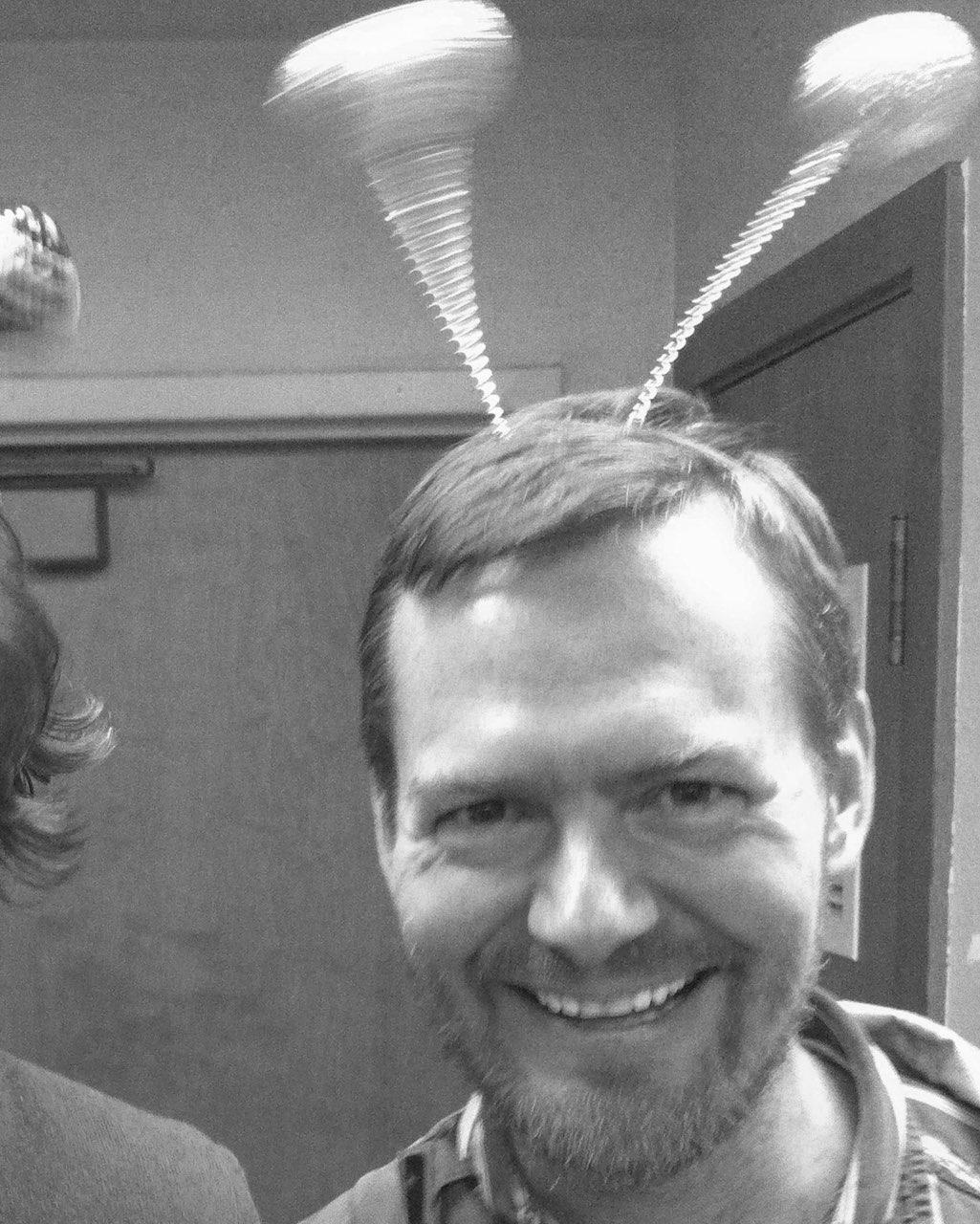}}]{Michael D. Ernst}
Michael Ernst is a Professor of Computer Science \& Engineering at the
University of Washington.
Ernst's research aims to make software more reliable, more secure, and
easier (and more fun!)\ to produce. His primary technical interests are in
software engineering, programming languages, type theory, security, program
analysis, bug prediction, testing, and verification. Ernst's research
combines strong theoretical foundations with realistic experimentation,
with an eye to changing the way that software developers work.
\end{IEEEbiography}

\begin{IEEEbiography}[{\includegraphics[width=1in,height=1.25in,clip,keepaspectratio]{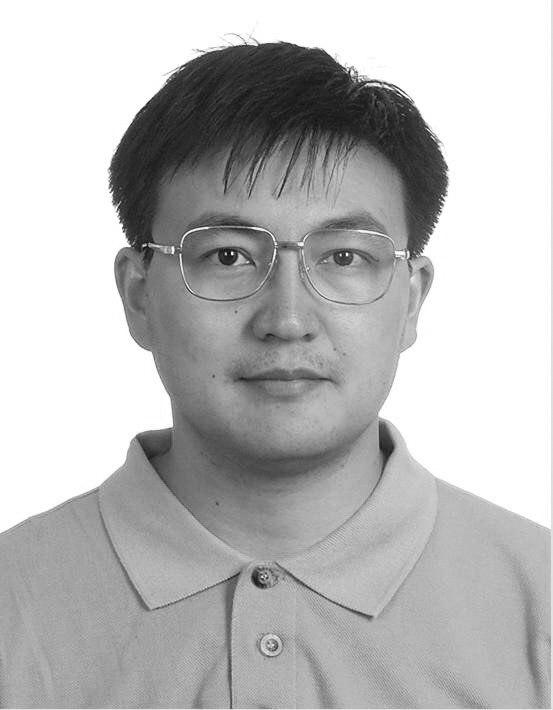}}]{Lu Zhang}
is a professor at the School of Electronics Engineering and Computer
Science, Peking University, P.R. China. He received both PhD and BSc
in Computer Science from Peking University in 2000 and 1995
respectively. He was a postdoctoral researcher in Oxford Brookes
University and University of Liverpool, UK\@. He served on the program
committees of many prestigious conferences, such FSE, OOPSLA, ISSTA,
and ASE\@. He was a program co-chair of SCAM2008 and a program co-chair
of ICSM17. He has been on the editorial boards of \textit{Journal of
Software Maintenance and Evolution: Research and Practice and Software
Testing, Verification and Reliability}. His current research interests
include software testing and analysis, program comprehension, software
maintenance and evolution,software reuse and component-based software
development, and service computing.
\end{IEEEbiography}



\end{document}